\documentstyle[prl,eqsecnum,aps,epsf]{revtex}
\begin{document}
\draft
\title{
\large \bf
Zero frequency divergence and gauge phase factor in the optical response 
theory
}
\author{Minzhong Xu}
\address{
Department of Chemistry, New York University, New York, NY 10003, U.S.A.
        }
\author{Xin Sun}
\address{ 
Department of Physics, Fudan University, and National Laboratory of Infrared
Physics, Shanghai 200433, P.R. China
}
\date{\today}
\maketitle
\bigskip
\begin{abstract}
The static current-current correlation leads to the definitional zero 
frequency divergence (ZFD) in the optical susceptibilities.
Previous computations have shown nonequivalent results between 
two gauges (${\bf p\cdot A}$ and ${\bf E \cdot r}$) 
under the exact same unperturbed wave functions.
We reveal that those problems are caused by the improper treatment of
the time-dependent gauge phase factor in the optical response theory. 
The gauge phase factor, which is conventionally ignored by the theory, is 
important in solving ZFD and obtaining the equivalent results between these 
two gauges. The Hamiltonians with these two gauges are not necessary
equivalent unless the gauge phase factor is properly considered in the
wavefunctions. Both 
Su-Shrieffer-Heeger (SSH) and Takayama-Lin-Liu-Maki (TLM) models of 
trans-polyacetylene serve as our illustrative examples to study the linear 
susceptibility $\chi^{(1)}$ through both current-current and dipole-dipole 
correlations. Previous improper results of the $\chi^{(1)}$ calculations and 
distribution functions with both gauges are discussed. The importance of 
gauge phase factor to solve the ZFD problem is emphasized based on SSH and TLM 
models. As a conclusion, the reason why dipole-dipole correlation favors over 
current-current correlation in the practical computations is explained.\\
\end{abstract}


\section{introduction}
The static current-current correlation ($J_0J_0$)\cite{wwu,mahan1,butcher}
is widely applied in the the optical response theory for many decades, both in 
the definition of linear susceptibility $\chi^{(1)}$ or conductivity 
$\sigma^{(1)},$\cite{mahan1} and in the definition of the nonlinear optical 
susceptibilities such as $\chi^{(n)}$, where $n \geq 2$.\cite{wwu,butcher}

Within the semiclassical theory of radiation, which is also literaturely
emphasized by Mahan,\cite{mahan1,mahan2} Bucher and Cotter,\cite{butcher}
Bloembergen,\cite{bloembergen} Shen,\cite{shen} and Mukamel,\cite{mukamel} the 
electric field are treated classically and propagation of electromagnetic waves
in a medium is governed by Maxwell's equations, that is, the electric field 
${\bf E}({\bf r},t)$ at some specific position ${\bf r}$ and time $t$ can be 
described as:\cite{note0}
\begin{eqnarray}
{\bf E}({\bf r}, t)={\bf E_0} e^{i{\bf k\cdot r}-i\omega t},
\label{e}
\end{eqnarray}
where ${\bf E_0}$ is the amplitude,  ${\bf k}$ and $\omega$ are the 
wave vector and frequency.
Then, the $\chi^{(n)}$ in a unit 
volumn $v$ under the static current-current ($J_0J_0$) correlation is 
conventionally defined as follows:\cite{wwu,butcher}
\begin{eqnarray}
&\chi&^{(n)}(\Omega; \omega_{1}, \ldots, \omega_{n})=-\delta_{n1}
\frac{n^{(e)}e^2}{\epsilon_0 m \omega_1^2} \hat{I}+
\displaystyle \frac{\chi_{j_0j_0}^{(n)}
( \Omega; \omega_{1}, \ldots, \omega_{n})}
{\epsilon_0 i\Omega \omega_1 \cdots \omega_n},
\label{jj}
\end{eqnarray}
where $\displaystyle \Omega \equiv -\sum_{i=1}^{n} \omega_{i}$,
$n^{(e)}$ and $m$ are the electron density and electron mass,
$\epsilon_0$ the dielectric constant, $\hat{I}$
a unit dyadic, $\delta_{n,1}$ Kronecker symbol, and
\begin{eqnarray}
&\chi&^{(n)}_{j_0j_0}(\Omega; \omega_{1}, \ldots, \omega_{n})= \frac{1}{n!}
\left[ \frac{1}{\hbar} \right]^n \int  d{\bf r}_{1} \cdots d{\bf r}_{n}
\int dt_{1} \cdots dt_{n} \nonumber \\
& &\int d{\bf r} dt\, e^{-i {\bf k \cdot r}+ i \Omega t} \langle \hat{T}
{\bf \hat{J}_0} ({\bf r},t) {\bf \hat{J}_0}({\bf r}_{1},t_{1}) \cdots
{\bf \hat{J}_0} ({\bf r}_{n},t_{n}) \rangle,
\label{JJ}
\end{eqnarray}
with $\hat{T}$ is the time-ordering operator and ${\bf \hat{J}_0}$ is the static
current operator.\cite{note0}

If we choose the static dipole-dipole ($DD$) correlation, the 
$n$th-order susceptibilities will be obtained as 
follows:\cite{butcher,mahan2,bloembergen,shen,mukamel}
\begin{eqnarray}
&\chi&^{(n)}(\Omega; \omega_{1}, \ldots, \omega_{n})= \frac{1}{n!}
\left[ \frac{i}{\hbar} \right]^n \int  d{\bf r}_{1} \cdots d{\bf r}_{n}
\int dt_{1} \cdots dt_{n} \nonumber \\
& &\int d{\bf r} dt\, e^{-i {\bf k \cdot r}+ i \Omega t} \langle \hat{T}
\hat{{\bf D}} ({\bf r},t) \hat{{\bf D}}({\bf r}_{1},t_{1}) \cdots
\hat{{\bf D}} ({\bf r}_{n},t_{n}) \rangle,
\label{DD}
\end{eqnarray}
where $\hat{{\bf D}}$ is the static dipole operator.

It is commonly held that the apparrent zero frequency divergence (ZFD) 
through the static current experssion Eq.(\ref{jj}) is only a virtual problem, 
and the gauge ${\bf E \cdot r}$ and ${\bf p \cdot A}$ will give the exactly 
same results with the same unperturbed wave functions.\cite{bassani} 
In other words, Eq.(\ref{jj}) and Eq.(\ref{DD}) should reach the same 
result if one proceeds properly. In linear response theory, 
for the homogeneous and isotropic medium, Martin and Schwinger have shown that 
the ZFD in conductivity $\sigma^{(1)}$ through current expressions could
be cancelled by introducing 
diagmagnetic term (${\bf A}^2$ term).\cite{martin} The cancellation of ZFD in 
the linear conductivity by diagmagnetic term is also discussed by Mahan, 
Haug and Jauho in their famous books,\cite{mahan1,haug} with a careful 
consideration on the limitation sequence between ${\bf k}$ and 
$\omega$.\cite{note1} In solid state, for either full-filled or empty bands, 
Aspnes had shown the equivalence of two gauges (from Eq.(\ref{jj}) and 
Eq.(\ref{DD})) based on the assumption of the 
cancellation of a ZFD term Eq.(2.6) in his seminal work of $\chi^{(2)}$ 
computations.\cite{aspnes} These works strengthen the common feelings of the 
equivalence between two gauges under the static dipole or the static current 
expressions. And the issue of ZFD to be a virtual problem was usually borrowed
with limited justification.

Although the `virtual' property of this ZFD are literally emphasized 
by a lot of people, strictly speaking, as to our knowledge, the solution to
this ZFD problem is seldom directly obtained from $J_0J_0$ 
correlation [like Eq.(\ref{jj})] except Martin and Schwinger's original
proof.\cite{martin} As we can see in the calculation of $\chi^{(2)}$ in solid 
state, the ZFD term is conventionally isolated from the convergent term and 
then is discarded without careful direct check.\cite{aspnes} Historically, 
this ZFD problem is of no interest due to the following facts: 
(i). The gauge transformations seem to guarantee the equivalence of two gauges 
under the same set of wave functions; (ii). In transport 
theory,\cite{haug,rammer} the correct imaginary part of $J_0J_0$ correlation 
(Re[$\sigma^{(1)}(\omega)$]) still can be obtained and (iii) the ZFD problem 
(related to Im[$\sigma^{(1)}(\omega)$]) usually can be avoided by
applying the Kramers-Kronig (KK) relations on the imaginary part of $J_0J_0$ 
correlation.\cite{mahan1,haug,rammer,callaway} Thus the equivalence of two 
gauges becomes 
a well-accepted opinion in the community and the ZFD is widely considered
to be at most a complex technical problem.\cite{mahan1,haug}

Besides those assumptions made in the previous ZFD 
proof,\cite{mahan1,martin,haug,note1,aspnes} however, in the practical 
applications, there always exists some puzzles from the above common feelings. 
In a careful study on the gauge choice of two-photon 
transitions in 1s-2s transition of hydrogen atom,\cite{bassani} Bassani, 
Forney and Quattropani have found that by directly applying the exact same 
unperturbed wave functions, ${\bf E \cdot r}$ converges much faster than 
${\bf p \cdot A}$ by using a limitted number of discrete intermediate states;
Numerical computations also show a 50 \% difference of the transition rate 
between those gauges if we only included all discrete intermediate states. 
Thus, Bassani {\it et al.} draw the conclusion that ${\bf E \cdot r}$ is a 
better gauge. While the study of oscillator strength in a superlattice, by 
Peeters {\it et al.}, turns out that ${\bf E \cdot r}$ and ${\bf p \cdot A}$ is
 nonequivalent in the barrier region based on numerical 
computations.\cite{peeters} Thus, on the contrary, Peeters {\it et al.} 
concluded that position operator ${\bf r}$ in the 
solid state should be redefined and ${\bf p \cdot A}$ is much better.
A recent study on zinc-blende semiconductor\cite{khurgin} again brings the
questions on the equivalence role of ${\bf E \cdot r}$ and ${\bf p \cdot A}$
in representing the transition matrix of $\chi^{(2)}$ formula.\cite{note2}

Extensive studies on the optical properties of 
polymers\cite{wwu,maki,kivelson,bishop,gebhard} 
are based on the tight-binding approximate (TBA) models, such 
as Su-Shrieffer-Heeger (SSH) model\cite{ssh} and Takayama-Lin-Liu-Maki (TLM) 
model\cite{tlm} in weakly correlated systems, Hubbard and 
Pariser-Parr-Pople (PPP) models in strongly correlated systems. Those 
models drastically reduce the complicity of the systems and provide a 
reasonable way to reveal the actual physical insights of many-body systems. 
In providing the gauge invariance, a U(1) phase transformation has been
suggested\cite{gebhard,fradkin} in those models. Based on the static current
formula without diagmagnetic term, there will be ZFD problem in the linear 
conductivity $\sigma^{(1)}(\omega)$. However, this ZFD problem has not been 
pointed out clearly and obviously neglected by the previous 
works.\cite{maki,gebhard} To avoid ZFD in 
$\sigma^{(1)}$ and to obtain a convergent result, Batistic and Bishop suggest 
to subtract the term $\langle [j_0,j_0] \rangle(\omega=0)$,\cite{bishop}
which is supposed to be the diagmagnetic term directly derived from 
Hamiltonian. Unfortunately, as we will show in this paper that the diagmagnetic
term derived from U(1) transformation\cite{gebhard} can not directly return the 
expected $\langle [j_0,j_0] \rangle(\omega=0)$ term. Moreover, experimentally 
observed two-photon absorption peak\cite{kajzar,heeger1} in the $\chi^{(3)}$
spectrum of trans-polyacetylene has raised wide interest for the theoretical
explanation. From Eq.(\ref{jj}) based on TLM models, two-photon cusp was 
obtained analytically,\cite{wwu} but it has been criticized by
the others from the dipole formula approach and other physical 
concern.\cite{su1,su2,cwu1,cwu2,shuai,mazumdar,soos} Recently, a quite 
different analytical form of the $\chi^{(3)}$ spectrum\cite{mxu} is obtained
under $DD$ correlation from that under $J_0J_0$ correlation.\cite{wwu} 
Those above discrepancies indeed already cast some doubts on the rooted belief 
of equivalence between $J_0J_0$ and $DD$ correlations.

In this paper, we will reexamine the concepts of the gauge transformations and
directly show that the gauge phase factor, which is not sufficiently emphasized
previously and is conventionally ignored under the current-current correlation 
scheme,\cite{wwu,butcher} is actually very important in the optical response 
theory to solve this ZFD difficulty, and to recover the equivalent results 
between the two gauges (${\bf p \cdot A}$ and ${\bf r \cdot E}$). Therefore, 
the static current operator $\hat{J}_0$ is no longer suitable for considering 
the equivalence between two gauges, instead, we should 
include the induced field currents (IFC) which are introduced by the gauge 
phase factor. To illustrate the effect of gauge phase factor rather than the
abstract concepts, we only choose the linear response under the 
one-dimensional ($1d$) periodic TBA models like SSH and TLM models as our 
examples. Certainly the concepts of the gauge phase factor under the specific 
linear examples also can be expanded into the nonlinear optical response theory
and two-dimensional ($2d$) or three-dimensional ($3d$) cases.

The paper is organized as follows: In Sec.II, we will reexamine the 
concept of the gauge transformation and discuss the importance of the gauge
phase factor in the optical response theory. The problems caused by the 
ignorance of the gauge phase factor are discussed under a general scheme 
independent of the models. To give an intuitive picture, linear optical 
response under periodic tight-binding-models (TBA) such as both SSH and TLM
models is investigated in this paper, the Hamiltonian under $DD$ 
correlation (${\bf E \cdot r}$) is discussed (Sec.III.A), and we will 
study the linear susceptibility $\chi^{(1)}$ through $DD$ correlation in 
Sec.III.B. The qualitative different results by the conventional 
polarization operator $\hat{P}$ in optical response are discussed based on 
periodic models in Sec.III.C. $\chi^{(1)}$ under current-current correlation 
is discussed in Sec.IV. The SSH Hamiltonian of ${\bf p \cdot A}$ will be 
obtained in Sec.IV.A. Before applying the gauge phase factor, the ZFD problem 
of $\chi^{(1)}$ through $J_0J_0$ correlation is 
illustrated in Sec.IV.B, previous qualitatively different solutions of 
$\chi^{(1)}$ and the practical difficulties in this ZFD problem based on 
the models are also analyzed (Sec.IV.B). After applying the gauge phase factor
to the wave functions, a convergent result can be obtained and the ZFD problem 
will be solved (Sec.IV.C).
The conditions of equivalence between two gauges are discussed in Sec.V.A 
and the influence of gauge phase factor on the initial distribution 
function $f_n({\bf k})$ under two gauges are investigated (Sec.V.B). The 
reasons of some previous puzzles on the choice of the gauges are discussed 
in Sec.V.C. The conclusion 
emphasizing the implications of our work will be described in Sec.VI.

\section{gauge phase factor in gauge transformation}

Gauge transformation has already been well-understood
in the optical response theory.\cite{schiff,cohen} The equivalence between
two gauges are built up on the concept of gauge transformation.

If the electromagnetic field applied, the Sch\"{o}dinger equation is given by:
\begin{eqnarray}
i\hbar\displaystyle\frac{\partial}{\partial t}\psi({\bf r},t)=
\left[\displaystyle\frac{1}{2m}(\hat{\bf p}-q {\bf A})^2+V({\bf r})
+q\phi\right]\psi({\bf r},t),
\label{se}
\end{eqnarray}

where $\psi({\bf r},t)$ is the exact wave function at space position ${\bf r}$
and specific time $t$, $m$ is the particle mass, $q$ is the 
electrical charge, $V({\bf r})$ is the potential, ${\bf A}$ and $\phi$ as 
vector and scalor potential 
correspondingly, under the following transformation:\\
\begin{equation}
\left \{
\begin{array}{l}
{\bf A} \to {\bf A'}={\bf A}+\nabla f({\bf r},t)\\
\phi \to \phi'=\phi-\displaystyle \frac{\partial}{\partial t} 
f({\bf r},t),
\end{array}
\right.
\label{gt1}
\end{equation}
where $f({\bf r},t)$ is arbitrary, and ${\bf A'}$ and $\phi'$ are new vector
and new scalor potentials after the transformation Eq.(\ref{gt1}). It could be 
shown\cite{cohen} that the form of the Sch\"{o}dinger equation will be 
exactly the same if the old wave function $\psi$ makes the following change
into the new exact wave function $\psi'$: 
\begin{eqnarray}
\psi \to \psi'=e^{iF_g({\bf r},t)}\psi=\hat{T}_G({\bf r},t)\psi,
\label{gt2}
\end{eqnarray}
where gauge phase factor $F_g({\bf r},t)$ is defined as:
\begin{eqnarray}
F_g({\bf r},t) \equiv \displaystyle \frac{q}{\hbar}f({\bf r},t).
\label{gf}
\end{eqnarray}

The above Eq.(\ref{gt1}) and Eq.(\ref{gt2}) are called gauge 
transformation (or $U(1)$ transformation\cite{fradkin}).

Long-wavelength approximation\cite{mahan1,butcher} is used in this paper, that
is, the ${\bf k}=0$ in Eq.(\ref{e}), and the electric
field ${\bf E}$ is described as ${\bf E}={\bf E_0} e^{-i\omega t}$.

If we consider the following initial scalor and vector potentials
under ${\bf E \cdot r}$ gauge:
\begin{eqnarray}
{\bf A}=0, \phi=-{\bf E} \cdot r.
\label{er}
\end{eqnarray}
After chosing the gauge phase factor $F_g$ as
\begin{eqnarray}
F_g=\displaystyle \frac{q{\bf E} \cdot r}{i\hbar\omega},
\label{fg}
\end{eqnarray}
by Eq.(\ref{gt1}), we obtain the new vector and new scalor potential under
${\bf p \cdot A}$ gauge as:
\begin{eqnarray}
{\bf A'}=\displaystyle\frac{\bf E}{i\omega}, \phi'=0.
\label{pa}
\end{eqnarray}
The connection between the old and new wave function is determined by 
Eq.(\ref{gt2}). 

Under perturbative schemes to study the optical response, 
conventionally people use the exact same set of unperturbed wave functions 
$\psi^0_n({\bf r},t)$ of Hamiltonian $\hat{H}_0$ (when ${\bf A}=0$ and 
${\psi=0}$ in Eq.(\ref{se})) to serve as our expansion basis for both 
${\bf E \cdot r}$ and ${\bf p \cdot A}$ 
gauges.\cite{butcher,bloembergen,bassani} However, we should
point out that the wave functions for both ${\bf E \cdot r}$ and 
${\bf p \cdot A}$ gauges (before and after gauge transformation)
should also be restricted by the gauge phase factor $F_g$ from 
Eq.(\ref{gt2}), therefore two basis sets for both gauges are {\bf not}
the exact same unperturbated wave functions $\psi^0_n({\bf r},t)$, but 
are different by the gauge phase factor $F_g$. And the Hamiltonian under
two gauges(${\bf E\cdot r}$ and ${\bf p \cdot A}$) are not necessary 
equivalent if they are treated independently and are isolated from the 
connection between the wave functions under the two gauges. Unfortunately, 
this crucial point has not been clearly illustrated and obviously missed
by the perturbated schemes works.\cite{butcher,bloembergen,shen} Especially 
under current-current correlation scheme, the gauge phase factor's 
contribution is obviously ignored 
and $A^2(t)$ term is considered as no any physical meanings.\cite{note3} 
Thus the current-current correlation is conventionally reduced into the 
$J_0J_0$ formula such as Eq.(\ref{jj}), and the equivalence 
between current-current and dipole-dipole correlations is usually considered 
as $J_0J_0$ and $DD$ correlations under the exact same basis of 
unperturbed wave functions.\cite{wwu,butcher,bloembergen,bassani,aspnes,peeters}

An elegant review by Langhoff, Epstein and Karplus covered the topics of 
time-dependent perturbative theory,\cite{langhoff} they have sharply pointed 
out that the time-dependent phase in wave functions is very essential and the
improper treatment of time-dependent phase will cause secular divergence in 
time-dependent perturbations. In field theory, it is also well-understood that
the improper treatment of the phase factor will cause divergence.\cite{mahan1}
Since the gauge phase factor Eq.(\ref{fg}) is obviously time-dependent, 
neglecting this phase factor will cause the ZFD in the susceptibility
computations, as the examples we will show in the following sections. 

\section{linear response by dipole-dipole correlation}

Having appreciated the importance of the gauge phase factor, we choose the 
following single electron periodic models -- SSH (or H\"{u}ckel) model and
TLM model as our examples provided the following reasons: (i). Those 
periodic models are widely applied in the polymer theory in the 80s and early
90s; Remarkable results have been obtained;\cite{heeger2} (ii). The optical 
susceptibilities obtained from those models can be analytically solvable and 
be compared with the previous results.\cite{maki,gebhard}
(iii). In both the SSH and the TLM models, Peirels instability\cite{heeger2} 
leads to the semiconductor property of two band structure -- with valence band 
full-filled and conduction band empty. It is very obvious from the physical
point of view, as the freqency of the electrical field goes $0$ (reaches
the static electric field), the linear conductivity $\sigma^{(1)}$ and the 
linear susceptibility $\chi^{(1)}$ will not be reduced to Drude
formula as in the metals,\cite{rammer,ferry} and will not cause ZFD problem.
In this section, we will first discuss $\chi^{(1)}$ and $\sigma^{(1)}$ of 
infinite chains (with number of $(CH)$ unit $N$ goes to infinity) under 
the $DD$ correlation. 

\subsection{SSH Hamiltonian}
Based on periodic TBA, The SSH Hamiltonian\cite{ssh} is given by:
\begin{eqnarray}
H_{SSH}=-\sum_{l,s} \left[ t_0+(-1)^l \frac{\Delta}{2} \right]
(\hat{C}_{l+1,s}^{\dag}\hat{C}_{l,s}^{}+\hat{C}_{l,s}^{\dag}\hat{C}_{l+1,s})^{},
\label{Hssh}
\end{eqnarray}
where $t_0$ is the transfer integral between the nearest-neighbor sites,
$\Delta$ is the gap parameter and $\hat{C}_{l,s}^{\dag}(\hat{C}_{l,s})$
creates(annihilates) an $\pi$ electron at site $l$ with spin $s$. In
continuum limitation, above SSH model will give the TLM model.\cite{tlm}
For the SSH model, each site is occupied by one electron. 

If we want to include the electron-photon interaction ${\bf E \cdot r}$ 
directly from the polarization operator $\hat{\bf P}$, where
\begin{eqnarray}
\hat{P}= \sum_{l} R_{l}\hat{C}^{\dagger}_{l} \hat{C}_{l},
\label{P}
\end{eqnarray}
and  
\begin{eqnarray}
R_l=la+(-1)^l u
\label{rl}
\end{eqnarray}
is the site $l$ position with the lattice constant $a$ and dimerized constant 
u,\cite{ssh}
we will face the problem of ill-definition of $\hat{\bf P}$ in periodic 
systems.\cite{aspnes,peeters,gebhard} To solve this problem, we should 
consider the imposed periodic condition of the position operator 
${\bf r}$.\cite{callaway,peeters} Expressing position operator
${\bf r}$ under the Bloch states $|n, {\bf k}> = u_{n, {\bf k}}
({\bf r})e^{i{\bf k \cdot r}}$, where $u_{n, {\bf k}}({\bf r})$ is the periodic
function under the translation of lattice vector,\cite{callaway} we will be
able to satisfy the periodic condition of ${\bf r}$ as follows:
\begin{eqnarray}
{\bf r}_{n {\bf k}, n' {\bf k'}}= i \delta_{n,n'}{\bf \nabla_{k}}
\delta({\bf k}-{\bf k'}) + \Omega_{n,n'}({\bf k})\delta({\bf k}-{\bf k'}),
\label{r}
\end{eqnarray}
and
\begin{eqnarray}
\displaystyle \Omega_{n,n'}({\bf k})=\frac{i}{v}\int_{v}
u_{n,{\bf k}}^*({\bf r}){\bf \nabla_{k}} u_{n', {\bf k}}({\bf r}) d {\bf r},
\end{eqnarray}
where $v$ is unit cell volume.

We change Hamiltonian Eq.(\ref{Hssh}) into the momentum space by applying 
the following consecutive transformations:
\begin{equation}
\left \{
\begin{array}{l}
\displaystyle \hat{C}_{l_o,s}=\frac{1}{\sqrt{N}}
\sum_{-\frac{\pi}{2a}\le k \le \frac{\pi}{2a}} 
(\hat{C}^v_{k,s}+\hat{C}^c_{k,s})e^{ikR_{l_o}},\\
\\
\displaystyle  \hat{C}_{l_e,s}=\frac{1}{\sqrt{N}}
\sum_{-\frac{\pi}{2a}\le k \le \frac{\pi}{2a}} 
(\hat{C}^v_{k,s}-\hat{C}^c_{k,s})e^{ikR_{l_e}},
\end{array}
\right.
\label{trk1}
\end{equation}
and
\begin{equation}
\left \{
\begin{array}{l}
\displaystyle \hat{a}^v_{k,s}=-i\gamma_k\hat{C}^v_{k,s}+\xi_k\hat{C}^c_{k,s},\\
\\
\displaystyle \hat{a}^c_{k,s}=i\xi_k\hat{C}^v_{k,s}+\gamma_k\hat{C}^c_{k,s},
\end{array}
\right.
\label{trk2}
\end{equation}
with
\begin{equation}
\left \{
\begin{array}{l}
\displaystyle \gamma_k=\frac{1}{\sqrt{2}}\sqrt{1+\frac{2t_0 cos(ka)}
{\varepsilon(k)}},\\
\\
\displaystyle \xi_k=\frac{sgn(k)}{\sqrt{2}}\sqrt{1-\frac{2t_0 cos(ka)}
{\varepsilon(k)}},
\end{array}
\right.
\end{equation}
where 
\begin{eqnarray}
\varepsilon (k)= \sqrt{\left[ 2 t_0 cos(ka) \right]^2+\left[ \Delta sin(ka) 
\right]^2}
\label{ek}
\end{eqnarray}
and $R_{l_o}$ and $R_{l_e}$ are odd and even position defined by Eq.(\ref{rl}).
$\hat{a}^{\dag c}_{k,s}(t)$ and $\hat{a}^{\dag v}_{k,s}(t)$ are 
the excitations of electrons in the conduction band and valence
band with momentum $k$ and spin $s$.

If we choose the spinor description $\hat{\psi}_{k,s}^{\dag}(t)$=
$(\hat{a}^{\dag c}_{k,s}(t)$, $\hat{a}^{\dag v}_{k,s}(t))$, SSH Hamiltonian 
including ${\bf E \cdot r}$ in momentum space is described by:
\begin{eqnarray}
\hat{H}_{SSH}(k,t)= \hat{H}_0+\hat{H}_{\bf E \cdot r},
\end{eqnarray}
where
\begin{eqnarray}
\hat{H}_0=\sum_{-\frac{\pi}{2a}\le k\le\frac{\pi}{2a},s} 
\varepsilon(k) \hat{\psi}_{k,s}^{\dag}(t) \sigma_{3}
\hat{\psi}_{k,s}(t)
\label{Hsshk}
\end{eqnarray}
and 
\begin{eqnarray}
\hat{H}_{\bf E \cdot r}=- \hat{D} \cdot E_0 e^{i\omega t}.
\end{eqnarray}

By Eq.(\ref{r}), the dipole operator $\hat{D}$ could be obtained
as follows:
\begin{eqnarray}
\hat{D}= e \sum_{-\frac{\pi}{2a}\le k\le\frac{\pi}{2a},s}
(\beta(k)\, \hat{\psi}_{k,s}^{\dag}
\sigma_{2}\hat{\psi}_{k,s}
 +i \frac{\partial}{\partial k} \, \hat{\psi}_{k,s}^{\dag}\hat{\psi}_{k,s}),
\label{D}
\end{eqnarray}
where
\begin{eqnarray}
\beta(k)=-\displaystyle \frac{\Delta t_0 a }{ \varepsilon^2(k)}+u,
\end{eqnarray}
is the coefficient related to the interband transition between the conduction
and valence bands in a unit cell $2a$ and the second term in Eq.(\ref{D}) is 
related to
the intraband transition,\cite{aspnes} $e$ is the electric charge and
$\vec{\sigma}$ are the Pauli matrixes. $u$ is dimerized constant related to 
the lattice distortion.\cite{ssh}

\subsection{Linear response through {$\bf E \cdot r$}}
For the linear susceptibility, $\chi_{SSH}^{(1)}(\Omega, \omega_1)$ can be 
obtained from Eq.(\ref{DD}) and Eq.(\ref{D}):
\begin{eqnarray}
\chi_{SSH}^{(1)}(-\omega_1, \omega_1)=2\left[ \frac{i}{\hbar} \right]
e^2 \sum_{k} \int_{-\infty}^{\infty} Tr \Biggl\{
& &i\frac{\partial}{\partial k} \left[G(k,\omega) i \frac{\partial}{\partial k}
\left[G(k,\omega -\omega_1) \right]\right] \nonumber\\
&+&\beta(k) \sigma_{2} G(k,\omega) i \frac{\partial}{\partial k} \left[
G(k,\omega-\omega_1) \right] \nonumber\\
&+&i \frac{\partial}{\partial k} \left[ \beta(k) G(k,\omega) \sigma_2
G(k,\omega-\omega_1) \right] \nonumber\\
&+&\beta(k) \sigma_2 G(k,\omega) \beta(k) \sigma_2 G(k,\omega-\omega_1)
\Biggr\} \frac{d \omega}{2 \pi},
\label{ssh1}
\end{eqnarray}
where the Green function $G(k,\omega)$ is the fourier transformation of
$G(t-t')\equiv-i\langle\hat{T}\hat{\psi}(t)\hat{\psi}^{\dag}(t')\rangle$ 
as follows:
\begin{eqnarray}
\displaystyle G(k,\omega)=
\frac{\omega+\omega_k\sigma_3}{\omega^2-\omega^2_k+i\epsilon},
\label{green}
\end{eqnarray} 
with $\omega_k \equiv \varepsilon(k) / \hbar \text{ and } \epsilon \equiv 0^+$.

By Eq.(\ref{ssh1}), we have $\chi_{SSH}^{(1)}(\omega)$$\equiv$
$\chi_{SSH}^{(1)}(-\omega,\omega)$:
\begin{eqnarray}
\chi_{SSH}^{(1)}(\omega)
= \frac{e^2(2 t_0 a)}{2 \pi \Delta^2} \int_{1}^{\frac{1}{\delta}}
\frac{(1-\eta\delta x^2)^2d x}{ [(1-\delta^2 x^2)(x^2-1)]^{\frac{1}{2}} 
x^2 (x^2-z^2)},
\label{sshx1}
\end{eqnarray}
where $ x \equiv \hbar \omega_k / \Delta$, $z \equiv \hbar \omega /(2\Delta)$,
$\delta \equiv \Delta /(2 t_0)$ and relative distortion $\eta \equiv (2u)/a$.
Eq.(\ref{sshx1}) can be numerically integrated if one change 
$x \to x+i\epsilon$ in considering the life-time of the 
state.\cite{su1,su2,cwu1,cwu2} For polyacetylene, by choosing $t_0=2.5 eV$, 
$\Delta=0.9 eV$, $a=1.22 \AA$, $u=0.04\AA$ and 
$\epsilon \sim 0.03$,\cite{su1,su2,cwu1,cwu2} we have $\delta=0.18$ and
$\eta=0.07$. The value of $|\chi^{(1)}_{SSH}|$ with or without $\eta$ 
contribution are plotted in Fig.1. As we can clearly see from the graph,
the relative distortion $\eta$'s contribution is very small (about $1\%$).
We can see the `unklapp enhancement' peak at $z=1$ comparing with the 
peak $z=1/\delta$. 
Those results are also discussed in the previous works.\cite{gebhard}
\begin{figure}
\vskip -10pt
\centerline{
\epsfxsize=7cm \epsfbox{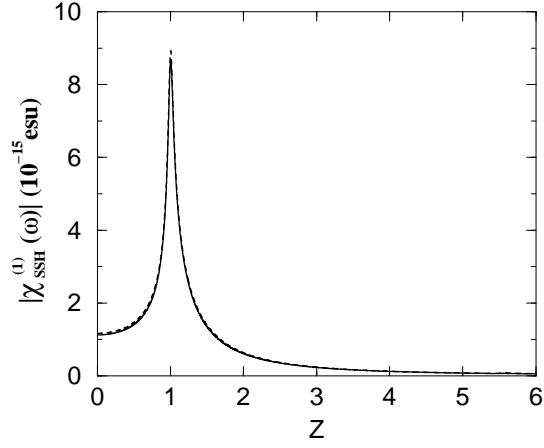}
}
\vskip -5pt
\caption{$|\chi_{SSH}^{(1)}(\omega)|$ with $z \equiv \hbar\omega/(2\Delta)$, 
for $\epsilon=0.03$, $\delta=0.18$ and for $\eta=0.07$ (solid line) or for
$\eta=0$ (dashed line).}
\end{figure}

If the continuum limitation is applied, that is, $\delta \to 0^+$, 
$\eta \to 0^+$, $\epsilon \to 0^+$ and $2t_0 a \to \hbar v_F $, the above 
integral Eq.(\ref{sshx1}) approaches the linear optical susceptibility
$\chi_{TLM}^{(1)}(\omega)$ under the TLM model\cite{tlm} as follows:
\begin{eqnarray}
\chi_{TLM}^{(1)}(\omega)=-\frac{e^2\hbar v_F}{2 \pi \Delta^2 z^2}(1-f(z)),
\label{x1}
\end{eqnarray}
where
\begin{equation}
f(z) \equiv \left \{
\begin{array}{lr}
\displaystyle  {\arcsin (z)\over z \sqrt{1-z^2}}  &(z^2<1),\\
\\
\displaystyle  -{\cosh^{-1} (z)\over z\sqrt{z^2-1}}+\displaystyle
{i\pi \over 2 z\sqrt{z^2-1}} &\ \ (z^2>1).
\end{array}
\right.
\label{fz}
\end{equation}
The conductivity $\sigma(\omega)$ given by $-i\omega \cdot \chi^{(1)}$. The
$Re[\sigma^{(1)}(\omega)]$ is the exact same as previous 
results.\cite{maki,gebhard}

The calculated $\chi^{(1)}_{TLM}$ and absolute value of $\chi^{(1)}_{TLM}$
are shown in Fig.2 and Fig.3.

\begin{figure}
\vskip -10pt
\centerline{
\epsfxsize=7cm \epsfbox{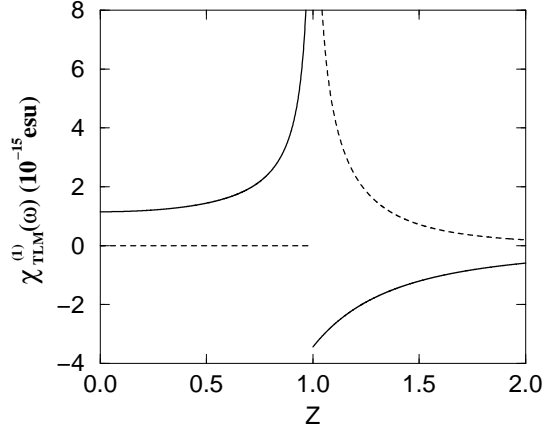}
}
\vskip 0pt
\caption{The real part (solid line) and the imaginary part (dashed line)
of $\chi_{TLM}^{(1)}(\omega)$ with $z \equiv \hbar\omega/(2\Delta)$.}
\end{figure}

\begin{figure}
\vskip -10pt
\centerline{
\epsfxsize=7cm \epsfbox{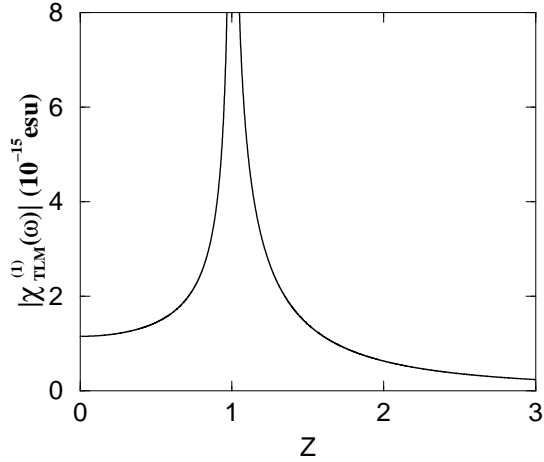}
}
\vskip -5pt
\caption{$|\chi_{TLM}^{(1)}(\omega)|$ with $z \equiv \hbar\omega/(2\Delta)$.}
\end{figure}

The above computations indeed are based on perturbative scheme with 
$\hat{\psi}^{\dag}_{k,s}(t)$ and $\hat{\psi}_{k,s}(t)$ as unperturbed creation 
and annihilation operators under Hamiltonian $\hat{H}_0$. $\left[\right.$
Eq.(\ref{Hsshk})$\left.\right]$ From the above figures and expressions, it is 
very obvious that the $DD$ correlation approach (or ${\bf E\cdot r}$)
will not have ZFD in $\chi^{(1)}$ and $\sigma^{(1)}$. The straightforward 
computations easily show that Eq.(\ref{sshx1}) and Eq.(\ref{x1}) obey the KK 
relation.  Those results are certainly reasonable under the physical picture.

\subsection{The qualitative different results through the polarization 
operator $\hat{P}$}
If the above dipole operator $\hat{D}$ is replaced by the 
polarization operator $\hat{P}$ defined by Eq.(\ref{P}), which are extensively 
used in the models,\cite{mukamel} we will obtain a different result from 
Eq.(\ref{x1}). As pointed in the 
literatures,\cite{aspnes,callaway,peeters,gebhard} the polarization
operator $\hat{P}$ is sensitive to boundary conditions and singular in the
thermodynamic limit. We will demonstrate this issue based on $\hat{P}$ in SSH 
model. Although the original point of position $R_l$ can be arbitrary, we can 
pick up the
chain region $l$ from $1$ to $N$ units to do a simple test. 
By Eq.(\ref{P}), Eq.(\ref{trk1})-Eq.(\ref{ek}), we obtain the 
unit polarization $\hat{P}^{unit}(k)$ from the total polarization 
$\hat{P}^{total}(k)$ in the momentum space: 
\begin{eqnarray}
\hat{P}^{unit}(k)=\lim_{N\to\infty}\displaystyle\frac{\hat{P}^{total}(k)}{N}=
\frac{ea}{2}\sum_{k,s} \psi^{\dag}_{k,s}\sigma_2\psi_{k,s}
\end{eqnarray}
We find that $\hat{P}^{unit}(k)$ contains no intraband transition like 2nd term 
in Eq.(\ref{D}). Applying a similar computation as $DD$ correlation,
we can obtain linear susceptibility $\chi^{(1)}_{SSH_P}$ based
on $\hat{P}^{unit}(k)$:
\begin{eqnarray}
\chi_{SSH_P}^{(1)}(\omega)
&=& \frac{e^2(2 t_0 a)}{2 \pi (2t_0)^2} \Biggl\{ 
\int_{1}^{\frac{1}{\delta}} \frac{d x}{[(1-\delta^2 x^2)(x^2-1)]^{\frac{1}{2}}}
+\int_{1}^{\frac{1}{\delta}} 
\frac{z^2dx}{[(1-\delta^2 x^2)(x^2-1)]^{\frac{1}{2}}(x^2-z^2)} \Biggr\} 
\nonumber \\
&=&\frac{e^2(2t_0a)\delta^2}{2\pi\Delta^2} \Biggl\{
F(\frac{\pi}{2},\sqrt{1-\delta^2})+\int_{1}^{\frac{1}{\delta}} 
\frac{z^2dx}{[(1-\delta^2 x^2)(x^2-1)]^{\frac{1}{2}}(x^2-z^2)} \Biggr\},
\label{sshx1p}
\end{eqnarray}
where $F(\pi/2,\sqrt{1-\delta^2})$ is the complete elliptic integrals
of the first kind. Obviously, the 1st term in the above Eq.(\ref{sshx1p}) is 
a constant and it is logarithm singular if $\delta\to0^+$! This result obeys 
the previous conclusions obtained in solid 
states,\cite{aspnes,callaway,peeters,gebhard} showing 
that the polarization operator's failure in optical susceptibility's study. 
We choose the same parameter as in Sec.III.B to plot spectrum of 
$|\chi_{SSH_P}^{(1)}|$, the comparison curves plotted
in Fig.4 showing that the results computed from $\hat{D}$ and $\hat{P}$ are
actually different both in the spectrum and in the magnitude. We find
another peak at $z=1/\delta$ corresponding to the transition from the bottom
of valence band to the top of conduction band is greater than the peak at
$z=1$. The 
`unklapp enhancement' disappears although the same position of resonant 
peak still can be predictable based on $\hat{P}$. Those differences show that
the electron transition from all the other sites can not be neglectable
especially in the periodic models. And conclusions obtained in $\chi^{(1)}$ 
on the dependence of gap parameter $\Delta$ are qualitatively different 
for an infinity chain between $\hat{D}$ and $\hat{P}$.\cite{mukamel1} 
Therefore there are some nonconsistencies between the polarization operator 
$\hat{P}$ and the dipole operator $\hat{D}$ at least in the linear 
response theory.
\begin{figure}
\vskip -10pt
\centerline{
\epsfxsize=7cm \epsfbox{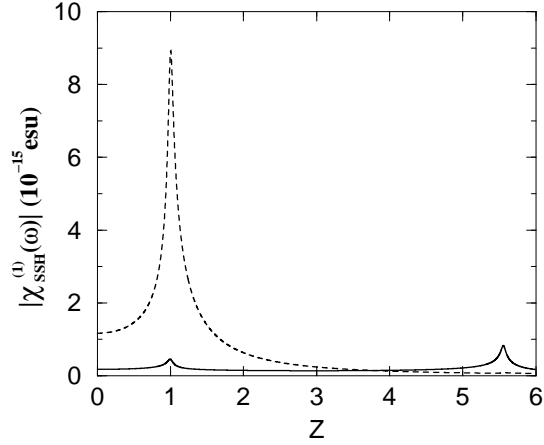}
}
\vskip 0pt
\caption{$|\chi_{SSH_P}^{(1)}(\omega)|$ under polarization operator $\hat{P}$ 
(solid line) versus $|\chi_{SSH}^{(1)}(\omega)|$ under dipole operator 
$\hat{D}$ (dashed line) with $z \equiv \hbar\omega/(2\Delta)$ and $\eta=0$. 
There is also a magnitude difference of $10^2$ between $\hat{D}$ and $\hat{P}$ 
besides the qualitative differences in the spectrum.}
\end{figure}

\section{linear response through current-current correlation}
\subsection{SSH Hamiltonian in the vector potential form}
The tight-binding Hamiltonian under ${\bf p \cdot A}$ form
should be invariant under gauge transformation (Eq.(\ref{gt1}) and 
Eq.(\ref{gt2})).\cite{gebhard,fradkin} If we change the phase of the 
one-particle wave function in tight-binding approximation:
\begin{eqnarray}
\hat{C}'_s({\bf r})=e^{i\theta}\hat{C}_s({\bf r}).
\label{wt}
\end{eqnarray}
we must modify the kinetic energy term according to the unperturbed 
Hamiltonian $\hat{H}_0$ in the Wannier functions basis as follows:
\begin{eqnarray}
H_0&\equiv&\sum_{s,{\bf r},{\bf r'}}t({\bf r}-{\bf r'})(C_s^{\dag}({\bf r})
C_s({\bf r'})+C_s^{\dag}({\bf r'})C_s({\bf r}))\\
&\to& \sum_{s,{\bf r},{\bf r'}}t({\bf r}-{\bf r'}) \left[ C_s^{'\dag}({\bf r})
e^{-iq\int^{\bf r'}_{\bf r}d{\bf x \cdot A(x)}/\hbar}C'_s({\bf r'}) \right.
\nonumber \\
& &+\left .C_s^{'\dag}({\bf r'})e^{iq\int^{\bf r'}_{\bf r}d{\bf x \cdot A(x)}
/\hbar} C'_s({\bf r})
\right ],
\label{HTBA}
\end{eqnarray}
where $t({\bf r}-{\bf r'})$ is the hopping from position ${\bf r}$ to 
${\bf r'}$, $C^{\dag}_s({\bf r})$ creates an electron at site ${\bf r}$ with
spin $s$, $q$ is the particle charge.
The above transformation is also known as Peierls substitution.\cite{gebhard}
The above form Eq.(\ref{HTBA}) has some kind of general meanings in TBA and
can be frequently seen in the  theoretical works.\cite{kirchner}

If the function $f({\bf r},t)$ is arbitrary in Eq.(\ref{gt1}), it is easy to 
verify that the above TBA Hamiltonian (\ref{HTBA}) is an invariant if the 
local phase change according to Eq.(\ref{wt}) or Eq.(\ref{gt2}):
\begin{eqnarray}
\theta({\bf r},t)\equiv \displaystyle \frac{e}{\hbar}f({\bf r},t).
\label{phase}
\end{eqnarray}

As a specific example, the SSH Hamiltonian with the vector potential
$A$ should be as follows (we change ${\bf A}$ into $A$ since it is 
$1d$ case):\cite{note4}
\begin{eqnarray}
H_{SSH}(A)&=&-\sum_{l,s} \left[ t_0+(-1)^l\frac{\Delta}{2}\right]
(\hat{C'}_{l+1,s}^{\dag}e^{-ieA(R_l-R_{l+1})/\hbar}\hat{C'}_{l,s}\nonumber \\
&+&\hat{C'}_{l,s}^{\dag}e^{ieA(R_l-R_{l+1})/\hbar}\hat{C'}_{l+1,s}),
\label{SSHA}
\end{eqnarray}

\subsection{result without gauge phase factor}
Under the assumption of the gauge phase factor (\ref{phase}) as no physical
meanings,\cite{butcher} we ignore the phase $\theta$ and treat the 
creation and annihilation operators $\{\hat{C'}^{\dag}_l\}$ and $\{\hat{C'}_l\}$
the same as unperturbed creation and annihilation operators 
$\{\hat{C}^{\dag}_l\}$ and $\{\hat{C}_l\}$\cite{gebhard,fradkin} 
defined in Eq.(\ref{Hssh}).

The above ${\bf p \cdot A}$ Hamiltonian Eq.(\ref{SSHA}) can be expanded in the 
powers of the external vector potential $A$ and obtain the following:
\begin{eqnarray}
\hat{H}_{SSH}(A)=\hat{H}_0-\hat{J}_0A-\frac{1}{2}\hat{J}_1A^2+O(A^3),
\label{Hssha}
\end{eqnarray}
where $H_0$ is given by Eq.(\ref{Hssh}),
\begin{eqnarray}
\hat{J}_0=-\sum_{l,s} \displaystyle &i&\frac{e}{\hbar} 
\left[ t_0+(-1)^l \frac{\Delta}{2} \right] 
\left[ a-2(-1)^l u \right] \nonumber \\
& &(\hat{C}_{l+1,s}^{\dag}\hat{C}_{l,s}^{}-\hat{C}_{l,s}^{\dag}
\hat{C}_{l+1,s})^{}
\label{j0}
\end{eqnarray}
and
\begin{eqnarray}
\hat{J}_1=-\sum_{l,s} \displaystyle & &(\frac{e}{\hbar})^2
\left[ t_0+(-1)^l \frac{\Delta}{2} \right] 
\left[ a-2(-1)^l u \right]^2 \nonumber \\
& &(\hat{C}_{l+1,s}^{\dag}\hat{C}_{l,s}^{}+\hat{C}_{l,s}^{\dag}
\hat{C}_{l+1,s})^{},
\label{j1}
\end{eqnarray}

The current operator $\hat{J}$ is obtained from the following 
equation:
\begin{eqnarray}
\hat{J}=\displaystyle \frac{i}{\hbar}\left[ \hat{P}, \hat{H} \right].
\end{eqnarray}

From Eq.(\ref{P}) and Eq.(\ref{Hssha}), we obtain the current operator under
the SSH Hamiltonian as follows:
\begin{eqnarray}
\hat{J}_{SSH}=\hat{J}_0+\hat{J}_1A,
\label{J}
\end{eqnarray}
where $\hat{J}_0$ and $\hat{J}_1$ are defined by Eq.(\ref{j0}) and 
Eq.(\ref{j1}).

Similar to the computations in Sec.III, we transform the
Hamiltonian Eq.(\ref{Hssha}) and the current operators Eq.(\ref{j0}) and 
Eq.(\ref{j1}) into the momentum space. By applying Eq.(\ref{trk1}) and 
Eq.(\ref{trk2}), we can obtain the following:
\begin{eqnarray}
\hat{J}_0(k)=\displaystyle \frac{ea}{\hbar} \sum_{k,s} \left[
A_0(k)\hat{\psi}_{k,s}^{\dag}(t)\sigma_3\hat{\psi}_{k,s}(t)
+B_0(k)\hat{\psi}_{k,s}^{\dag}(t)\sigma_1\hat{\psi}_{k,s}(t) \right]
\label{j0k}
\end{eqnarray}
and
\begin{eqnarray}
\hat{J}_1(k)=\displaystyle (\frac{ea}{\hbar})^2 \sum_{k,s} \left[
A_1(k)\hat{\psi}_{k,s}^{\dag}(t)\sigma_3\hat{\psi}_{k,s}(t)
+B_1(k)\hat{\psi}_{k,s}^{\dag}(t)\sigma_1\hat{\psi}_{k,s}(t)\right]
\label{j1k}
\end{eqnarray}
with $A_0(k), B_0(k), A_1(k) \text{and } B_1(k)$ defined as follows:

\begin{equation}
\left \{
\begin{array}{l}
A_0(k)=-\displaystyle\frac{(2t_0)^2(1-\delta^2)sin(2ka)}{2\varepsilon (k)},\\
\\
B_0(k)=-\displaystyle\frac{2t_0\Delta}{\varepsilon (k)}
+\eta \varepsilon (k)
\end{array}
\right.
\label{AB0}
\end{equation}
and
\begin{equation}
\left \{
\begin{array}{l}
A_1(k)=(1+\eta^2)\varepsilon (k)-
\displaystyle\frac{4t_0\Delta\eta}{\varepsilon (k)},\\
\\
B_1(k)=\displaystyle\frac{(2t_0)^2\eta(1-\delta^2)sin(2ka)}{\varepsilon (k)}
\end{array}
\right. 
\end{equation}
$\eta$, $\delta$ and $\varepsilon (k)$ in the above equations are 
defined the same as in the Sec.III. 

Applying Eq.(\ref{jj}) and Eq.(\ref{JJ}), and considering the 
diagmagnetic current $\hat{J}_1(k)A$ obtained from the above SSH Hamiltonian 
Eq.(\ref{Hssha}), we have the linear susceptibility under $J_0J_0$ correlation
as follows:\\
\begin{eqnarray}
{\chi'}_{SSH}^{(1)}(-\omega_1, \omega_1)=\displaystyle 
\frac{{\chi'}^{(1)}_{j_0j_0}(-\omega_1, \omega_1)}{-i\omega_1^2}
+\frac{{\chi'}^{(1)}_{j_1}(-\omega_1, \omega_1)}{-i\omega_1^2},
\label{sshjj}
\end{eqnarray}
with
\begin{eqnarray}
{\chi'}^{(1)}_{j_0j_0}(-\omega_1, \omega_1)=2\left[ \frac{1}{\hbar} \right]
(\frac{ea}{\hbar})^2 \sum_{k} \int_{-\infty}^{\infty} Tr \Biggl\{
& &A_0(k)\sigma_3G(k,\omega)A_0(k)\sigma_3G(k,\omega-\omega_1)\nonumber\\
&+&A_0(k)\sigma_3G(k,\omega)B_0(k)\sigma_1G(k,\omega-\omega_1)\nonumber\\
&+&B_0(k)\sigma_1G(k,\omega)A_0(k)\sigma_3G(k,\omega-\omega_1)\nonumber\\
&+&B_0(k)\sigma_1G(k,\omega)B_0(k)\sigma_1G(k,\omega-\omega_1)
\Biggr\} \frac{d \omega}{2 \pi},
\end{eqnarray}
\begin{eqnarray}
{\chi'}^{(1)}_{j_1}(-\omega_1, \omega_1)=-2(\frac{ea}{\hbar})^2
\sum_{k} \int_{-\infty}^{\infty} Tr \Biggl\{
& &A_1(k)\sigma_3G(k,\omega-\omega_1) \nonumber\\
&+&B_1(k)\sigma_1G(k,\omega-\omega_1)
\Biggr\} \frac{d \omega}{2 \pi},
\end{eqnarray}
where Green function $G(k,\omega)$ is defined by Eq.(\ref{green}). Following
the straightforward computations, we can obtain the following:
\begin{eqnarray}
{\chi'}_{j_0j_0}^{(1)}(\omega)
= -2i\frac{e^2(2 t_0 a)}{\pi \hbar^2} \int_{1}^{\frac{1}{\delta}}
\frac{(1-\eta\delta x^2)^2d x}{ [(1-\delta^2 x^2)(x^2-1)]^{\frac{1}{2}}
(x^2-z^2)},
\label{j0j0}
\end{eqnarray}
and
\begin{eqnarray}
{\chi'}_{j_1}^{(1)}(\omega)
= 2i\frac{e^2(2 t_0 a)}{\pi \hbar^2} \int_{1}^{\frac{1}{\delta}}
\frac{[(1+\eta^2)\delta^2x^2-2\eta\delta] dx}
{[(1-\delta^2 x^2)(x^2-1)]^{\frac{1}{2}}}.
\label{j1j1}
\end{eqnarray}
Eq.(\ref{j1j1}) is similar to the first term in Eq.(\ref{sshx1p}) and
is a constant independent of $z$. Eq.(\ref{j1j1}) obviously is not the
term $\langle [j_0,j_0] \rangle(\omega=0)$ suggested by Batistic and 
Bishop.\cite{bishop}

By Eq.(\ref{sshjj}), Eq.(\ref{j0j0}) and Eq.(\ref{j1j1}), we obtain:
\begin{eqnarray}
{\chi'}_{SSH}^{(1)}(\omega)
= \frac{e^2(2 t_0 a)}{2 \pi \Delta^2 z^2} \Biggl\{ \int_{1}^{\frac{1}{\delta}}
\frac{(1-\eta\delta x^2)^2d x}{ [(1-\delta^2 x^2)(x^2-1)]^{\frac{1}{2}}
(x^2-z^2)} - \int_{1}^{\frac{1}{\delta}}
\frac{[(1+\eta^2)\delta^2 x^2-2\eta\delta]d x}
{[(1-\delta^2 x^2)(x^2-1)]^{\frac{1}{2}}} \Biggr \},
\label{jsshx1}
\end{eqnarray}
where $x$,$\Delta$, $z$,$\delta$ and $\eta$ are all the same defined as in
Eq.(\ref{sshx1}).

If the continuum limitation (same as in Sec.III) is applied,
we will find that the contribution (Eq.(\ref{j1j1})) from the diagmagnetic 
term ($J_1$ term) disappears. We will obtain the following 
susceptibility under TLM models:
\begin{eqnarray}
{\chi'}_{TLM}^{(1)}(\omega)=\frac{e^2\hbar v_F}{2 \pi \Delta^2 z^2}f(z),
\label{jx1}
\end{eqnarray}
where $f(z)$ is defined by Eq.(\ref{fz}).

We plot $|{\chi'}^{(1)}_{SSH}|$, ${\chi'}^{(1)}_{TLM}$ and 
$|{\chi'}^{(1)}_{TLM}|$ in Fig.5, Fig.6 and Fig.7 with the same parameter
as in Sec.III. From the analytical form Eq.(\ref{jx1}) and the Figures, 
we find that the first term 
in Eq.(\ref{x1}) disappears in this $J_0J_0$ formula, however, this important
feature has not been reported in the previous work.\cite{maki,gebhard}

\begin{figure}
\vskip -10pt
\centerline{
\epsfxsize=7cm \epsfbox{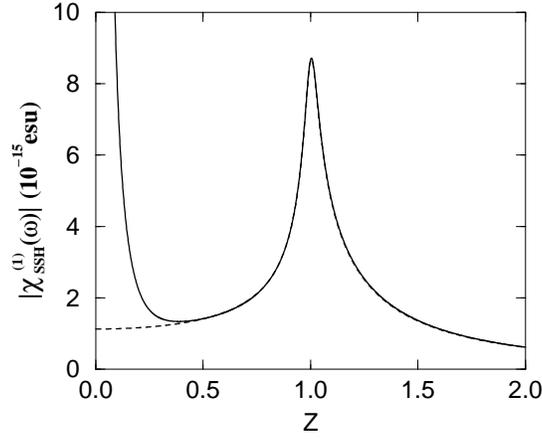}
}
\vskip -5pt
\caption{$|{\chi'}_{SSH}^{(1)}(\omega)|$ through $J_0J_0$
correlation (solid line) versus $|\chi_{SSH}^{(1)}(\omega)|$ through $DD$
correlation (dashed line) with $z \equiv \hbar\omega/(2\Delta)$, for
$\epsilon=0.03$, $\eta=0.07$ and $\delta=0.18$. $|{\chi'}_{SSH}^{(1)}(\omega)|$
obviously shows the ZFD if the gauge phase factor is not considered.}
\end{figure}

\begin{figure}
\vskip -10pt
\centerline{
\epsfxsize=7cm \epsfbox{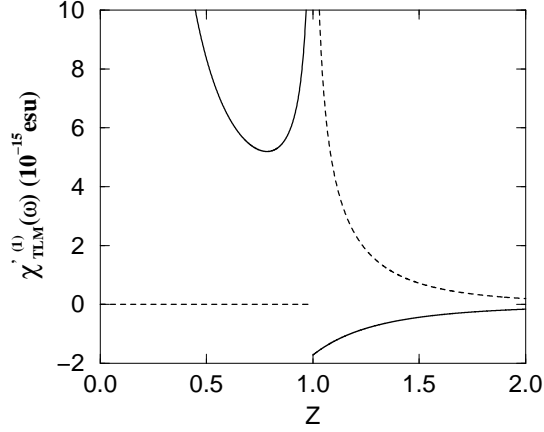}
}
\vskip 0pt
\caption{The real part (solid line) and the imaginary part (dashed line)
of ${\chi'}_{TLM}^{(1)}(\omega)$ with $z \equiv \hbar\omega/(2\Delta)$. It
shows the ZFD in real part when the gauge phase factor is not considered.}
\end{figure}

\begin{figure}
\vskip -10pt
\centerline{
\epsfxsize=7cm \epsfbox{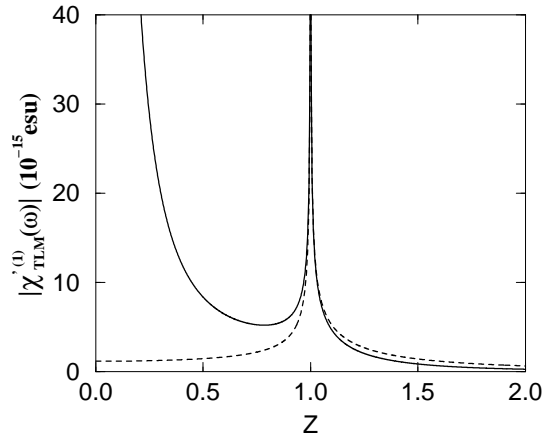}
}
\vskip -5pt
\caption{$|{\chi'}_{TLM}^{(1)}(\omega)|$ through $J_0J_0$ correlatoin
(solid line) compared with $|\chi_{TLM}^{(1)}(\omega)|$ through $DD$ 
correlation or $JJ$ correlation through gauge phase factor 
(dashed line) with $z \equiv \hbar\omega/(2\Delta)$. $|{\chi'}_{TLM}^{(1)}
(\omega)|$ shows ZFD.}
\end{figure}

Under TLM models, the missing diagmagnetic term can be understood directly 
form the Lagrangian.\cite{note5}
Obviously, Eq.(\ref{jsshx1}) and Eq.(\ref{jx1}) strongly diverge 
when $z\to 0$ for the real part in both Eq.(\ref{jsshx1}) and Eq.(\ref{jx1}).
The above results are certainly wrong since it does not follow the 
KK relation besides this ZFD problem. Careful comparisons are made between
Eq.(\ref{jsshx1}), Eq.(\ref{jx1}) and Eq.(\ref{sshx1}), Eq.(\ref{x1}), showing
that the correct imaginary parts obtained from the $J_0J_0$ correlation
Eq.(\ref{jj}) and Eq.(\ref{JJ}) are still maitained. It also shows that based 
on $J_0J_0$ correlation, the absorption part (related to imaginary part) will 
still be correct. This conclusion is not novel to the peoples working on the 
transport problem, where the Kubo formula based on $J_0J_0$ correlation is 
already applied as a common knowledge.\cite{mahan1} Peoples usually resort
KK relation to avoid real part ZFD difficulty.\cite{rammer}

But how to understand those difficulties (such as ZFD and violation of KK 
relation in Eq.(\ref{jsshx1}) and Eq.(\ref{jx1})) becomes a task.
The long wavelength approximation used in the above examples already
eliminate the possible ZFD caused by the limitation sequence between 
${\bf k}$ and $\omega$.\cite{note1} Someone maybe argue to solve 
the ZFD problem by including a diagmagnetic term, such as the effective 
mass $m^*$ and electron density $n_0$ assumption be made in this solid 
state problem.\cite{ferry,callaway} In fact, those assumptions are awkward 
because the parameters $m^*$ and $n_0$ in diagmagnetic term cannot be 
predictable in the models. For example,
in the above SSH and TLM models, $m^*$ and $n_0$ could be 
arbitrary.\cite{heeger2} In the TLM model, you can not include diagmagnetic 
term directly from Lagrangian.\cite{note5} For the SSH model, the diagmagnetic 
term can be included in Eq.(\ref{J}) from the TBA Hamiltonian Eq.(\ref{Hssha}),
but you still cannot solve this ZFD problem, as we clearly see from Fig.5 and 
Eq.(\ref{jsshx1}). Moreover, based on diagmagnetic terms to cancel ZFD, the 
assumption on the property of medium should be made in the 
proof.\cite{mahan1,martin,haug} 

The conventional way to treat this ZFD problem is to separate the divergent 
terms from the convergent term and directly throw it away with some possible 
physical explanations.\cite{aspnes} Fortunately, in linear response, neglect 
of diagmagnetic term does not cause so many troubles since the imaginary part 
of $J_0J_0$ correlation is still correct.\cite{mahan1,haug,rammer} It is not 
a big surprise that people have not taken this ZFD problem seriously although 
the doubts are always existing in those two gauges especially in the 
models.\cite{bassani,peeters,cohen}

\subsection{solving ZFD by gauge phase factor}
As we demonstrate above, the diagmagnetic term directly obtained from the 
Eq.(\ref{SSHA}) can not solve the ZFD problem. This ZFD problem, which is a 
conceptual problem as we have already pointed out in the Sec.II, is caused by 
the conventional careless treatment of the gauge phase factor in the optical 
response theory. In this part, we will demonstrate the same result of SSH 
and TLM models as $DD$ correlation after considering the contribution of 
gauge phase factor in $JJ$ correlation.

As we discussed in Sec.II, the new creation operator $\hat{C'}^{\dag}_l$ and 
annihilation operator $\hat{C'}_l$ in Eq.(\ref{SSHA}), should be differred by
the gauge phase factor from the unperturbed creation operator 
$\hat{C}^{\dag}_l$ and annihilation operators $\hat{C}_l$. Following
Eq.(\ref{gt2}), we obtain the following after the local phase factor is 
considered:\cite{note6}
\begin{eqnarray}
\hat{C'}_l=e^{ieAR_l/\hbar}\hat{C}_l
\end{eqnarray}

Thus by the above relation, the Hamiltonian in momentum space should make
the following change:
\begin{eqnarray}
\hat{H}(k)\to\hat{H}({\kappa}),
\end{eqnarray}
where
\begin{eqnarray}
\kappa=k+\frac{eA}{\hbar}.
\end{eqnarray}

For the SSH Hamiltonian, we have the following new Hamiltonian:
\begin{eqnarray}
H^{new}(k)=H_0^{new}(k)+H_1^{new}(k)A+O(A^2),
\end{eqnarray}
where $H_0^{new}(k)$ defined as the Eq.(\ref{Hsshk}) and
\begin{eqnarray}
H_1^{new}(k)=-\frac{ea}{\hbar}\sum_k B_0(k)
\hat{\psi}_{k,s}^{\dag}(t)\sigma_1\hat{\psi}_{k,s}(t).
\end{eqnarray}
 
The new current operator $\hat{J}^{new}(k)$ could be obtained from the
commutator equation $[\hat{D}(k),\hat{H}^{new}(k)]/(i\hbar)$ as
the following:
\begin{eqnarray}
\hat{J}^{new}(k)=\hat{J}^{new}_0(k)+\hat{J}^{new}_1(k)A+O(A^2),
\end{eqnarray}
where $\hat{J}^{new}_0(k)$ is exact the same as Eq.(\ref{j0k}) and
\begin{eqnarray}
\hat{J}_1^{new}(k)=\displaystyle (\frac{ea}{\hbar})^2 \sum_{k,s} \left[
A_1^{new}(k)\hat{\psi}_{k,s}^{\dag}(t)\sigma_3\hat{\psi}_{k,s}(t)
+B_1^{new}(k)\hat{\psi}_{k,s}^{\dag}(t)\sigma_1\hat{\psi}_{k,s}(t)\right]
\end{eqnarray}
where
\begin{equation}
\left \{
\begin{array}{l}
A_1^{new}(k)=\displaystyle\frac{B_0^2(k)}{\varepsilon (k)}\\
\\
B_1^{new}(k)=-A_0(k)\left[ \displaystyle\frac{2t_0\Delta}{\varepsilon^2(k)}
+\eta \right].
\end{array}
\right.
\end{equation}
where $A_0(k)$ and $B_0(k)$ are defined in Eq.(\ref{AB0}).
 
Thus, we find that after considering the gauge factor, the new current
operator $J_0^{new}(k)$ is the same as the static current $J_0(k)$, but 
$J_1^{new}(k)$ is different from the static current $J_1(k)$, we call those 
current differences between $J^{new}(k)$ and static currents $J_0(k)$, $J_1(k)$,
etc., as induced field currents (IFC) since they are introduced by the gauge 
field.

Through the evolution operator in interaction picture,\cite{mahan1} it is 
easy to derive the formula of the $\chi^{(1)}_{jj}$ as follows:
\begin{eqnarray}
\chi^{(1)}_{j^{new}j^{new}}(-\omega_1, \omega_1)=\displaystyle
\frac{\chi^{(1)}_{j_0^{new}h_1^{new}}(-\omega_1, \omega_1)
+\chi^{(1)}_{j_1^{new}}(-\omega_1, \omega_1)}
{-i\omega_1^2}
\label{jjnew}
\end{eqnarray}
where
\begin{eqnarray}
\chi^{(1)}_{j_0^{new}h_1^{new}}(-\omega_1, \omega_1)
=2\left[ \frac{1}{\hbar} \right]
(\frac{ea}{\hbar})^2 \sum_{k} \int_{-\infty}^{\infty} Tr \biggl\{
& &A_0(k)\sigma_3G(k,\omega)B_0(k)\sigma_1G(k,\omega-\omega_1)\nonumber\\
&+&B_0(k)\sigma_1G(k,\omega)B_0(k)\sigma_1G(k,\omega-\omega_1)
\biggr\} \frac{d \omega}{2 \pi},
\label{j0j0new}
\end{eqnarray}
\begin{eqnarray}
{\chi'}^{(1)}_{j_1^{new}}(-\omega_1, \omega_1)=-2(\frac{ea}{\hbar})^2
\sum_{k} \int_{-\infty}^{\infty} Tr \biggl\{
& &A_1^{new}(k)\sigma_3G(k,\omega-\omega_1) \nonumber\\
&+&B_1^{new}(k)\sigma_1G(k,\omega-\omega_1)
\biggr\} \frac{d \omega}{2 \pi},
\label{j1new}
\end{eqnarray}

Eq.(\ref{j0j0new}) gives exact the same result as Eq.(\ref{j0j0}),
which is computed through $J_0J_0$ correlation. While the contribution from 
$J_1^{new}$ in Eq.(\ref{j1new}) can be obtained as:
\begin{eqnarray}
{\chi'}^{(1)}_{j_1^{new}}(\omega)
= 2i\frac{e^2(2 t_0 a)}{\pi \hbar^2} \int_{1}^{\frac{1}{\delta}}
\frac{(1-\eta\delta x^2)^2d x}{ [(1-\delta^2 x^2)(x^2-1)]^{\frac{1}{2}}
x^2},
\label{j1j1new}
\end{eqnarray}

The above term (\ref{j1j1new}) actually is $\langle [j_0,j_0] \rangle(\omega=0)$
suggested by Batistic and Bishop.\cite{bishop}  

By Eq.(\ref{j0j0}) and Eq.(\ref{j1j1new}),  Eq.(\ref{jjnew}) returns 
the exact the same result as Eq.(\ref{sshx1}) and Eq.(\ref{x1}), which are
computed through $DD$ correlation.

\section{discussions} 
\subsection{equivalence condition between two gauges}
The above examples based on $J_0J_0$ and $DD$ correlations show the importance 
of the gauge phase factor $F_g$ in the optical response theory. 
The equivalence of the two gauges (${\bf E \cdot r} \text{and } 
{\bf p \cdot A}$) should not be based on the exact same sets of wave 
functions, but should be different by the gauge phase factor. This 
crucial point has never been clearly pointed out previously.
The conventional equivalence between the static current-current ($J_0J_0$) 
correlation and the static dipole-dipole ($DD$) correlation is not maintained,
instead, we should consider the induced field currents (IFCs) which is 
introduced by the gauge phase factor. For example, in the periodic models, we 
should do the phase shift from ${\bf k} \to {\bf\kappa}$ for the basis wave 
functions (or through creation and annihilation operators in models)  to obtain 
the new current operator $\hat{J}^{new}$ for linear case. Thus, the 
equivalence between current-current and dipole-dipole correlations should be 
understood as the situation when all IFCs are included. This paper's conclusion
that the time-dependent gauge phase factor $F_g$ can not be ignored under the 
perturbative 
scheme is also consistent with Langhoff {\it et al.}'s results.\cite{langhoff}

\subsection{Initial distribution function in two gauges}
In optical response, people are much more interested in the population of the 
states than the phase factor. Recently, femtochemistry 
experiments\cite{zewail} are able to reveal the phase factor's effect
by the vibrational modes of nucleis other than the population effect
(or distribution function) of states. To some extend, the phase factor's
influence on the optical response begin to raise people's interest.
Although our treatment of the optical susceptibilities is still under the 
frame of non-vibration nucleis, the gauge phase factor's effect in the
theory still can be seen from the above computations.

Another important feature caused by gauge phase factor $F_g$ is the influence 
on the initial distribution function $f_n({\bf k})$ between two gauges. If all 
particles are in the ground state $\psi_g^0$ of unperturbed Hamiltonian 
$\hat{H}_0$ when the electric field ${\bf E}$ is applied at time $t=0$, 
under the perturbative scheme, we can use the set of unperturbed wave functions 
$\{\psi_n^0\}$ with the initial distribution function 
\begin{eqnarray}
f_g({\bf k})=1 \quad \text{and} \quad f_n({\bf k})=0 \text{ for all the other} 
(n\ne g); 
\label{id}
\end{eqnarray}
Previously, we use the exact the same initial distribution function such like
Eq.(\ref{id}) for both gauges.

As we show in the previous sections, the exact wave functions 
between two gauges should be different by a time-dependent gauge phase factor.
Those conclusions are also kept under the perturbative scheme.
Thus, the initial distribution function in two different gauges are not
necessary the exact same as Eq.(\ref{id}) and should be carefully considered 
on the choice of the basis sets.

Specifically for ${\bf E\cdot r}$, we should choose unperturbed wave functions 
as the basis to avoid ZFD directly,\cite{note7} then the initial distribution 
function could be Eq.(\ref{id}) since the initial ground state is $\psi_g^0$. 
But for ${\bf p\cdot A}$, since the new ground state should be 
$e^{iF_g}\psi_g^0$ according to Eq.({\ref{gt2}), there are two ways to set the 
initial distribution function: (i)If we have already considered the gauge phase
factor in our new basis set, that is, we use $\{e^{iF_g}\psi_n^0\}$ as our 
basis, the distribution function is still as Eq.(\ref{id}); (ii)If we still use 
unperturbed wave function $\{\psi^0_n\}$ as our basis, we should project the 
initial wave function $e^{iF_g}\psi_g^0$ on the basis set $\{\psi_n^0\}$ 
instead of directly applying Eq.(\ref{id}) as our initial distribution function.
The latter way is much more complicated since the initial set of distribution 
function will be time-dependent. In the previous example we show, we use the 
first way under the local phase approximation.\cite{note6} Both ways to treat 
initial distribution function by ${\bf p\cdot A}$ are complicated, in this 
meaning, the gauge ${\bf E\cdot r}$ is much better. 
This conclusion has actually already been tested in the practical 
applications.\cite{mahan1,butcher,mahan2,bloembergen,shen,mukamel,bassani,haug,aspnes,su1,su2,cwu1,cwu2,shuai,mazumdar,soos,mxu}

\subsection{some previous puzzles}
In understanding those nonequivalent results between the two gauges, some 
literatures are emphasized by the others.\cite{note8} However, those 
explanations lacks the direct proof and the effect of the gauge phase 
factor is ignored. The above two subsections provide the qualitative
reasons to understand those nonequivalent puzzles between
the two gauges in the numerical computations.\cite{bassani,aspnes,mxu} 
Based on the same set of unperturbed wave functions,\cite{bassani} the 
computations of Bassani {\it et al.} shows that the ${\bf E\cdot r}$ converges 
much faster than ${\bf p \cdot A}$ reveal that the distribution Eq.(\ref{id}) 
is good for ${\bf E \cdot r}$ but not for ${\bf p \cdot A}$ case, which is 
certainly reasonable according to our discussions in Sec.V.B. 
The $\chi^{(2)}$ computation is a special case,\cite{aspnes,note3} since some 
terms could be $0$ if the symmetry is applied.\cite{cohen} Unfortunately, 
the nonequivalent results between $J_0J_0$ correlation and $DD$ correlation
are magnified in the $\chi^{(3)}$ 
computations. As an example, the spectrum of $\chi^{(3)}$ of polyacetylene
shows a different spectrum between two gauges.\cite{wwu,mxu} Those 
difference could be understood qualitatively through the gauge phase factor.

\section{conclusions}
From the computations and discussions above, we concluded that
Eq.(\ref{jj}) based on the static current is improper and leads ZFD
problem, thus, IFCs generated from the gauge phase factor should be included
to solve this difficulty. Generally speaking, the Hamiltonians under two gauges
are not necessary equivalent unless the gauge phase factor is properly
considered through the wave functions. Because the choice problems in the 
initial distribution function and the basis sets under two gauges (Sec.V.B), it
will be very tedious to do the perturbative
computations based on the ${\bf p \cdot A}$ than ${\bf E \cdot r}$. If the 
careful computation is made based on the concept of gauge phase factor, both 
gauges will lead to the equivalent result. Although our computations are 
chiefly based on a $1d$ periodic model, it could be easily seen that the chief 
conclusions of this paper can be expanded in the $2d$, $3d$ and other systems
based on the general illustrations in Sec.II.

\acknowledgments 
One of the author (M. Xu) would thank for Professor Y. Takahashi for intuitive
discussions on some concepts in the filed theory during his visit to Fudan 
University in 1992. Very helpful discussions with Professor J.L. Birman, 
Professor H.L. Cui, Professor Y.R. Shen, Professor Z.G. Soos, Professor 
C.Q. Wu and Dr. Z.G. Yu are acknowledged. This work was supported by Chemistry 
Department, New York University, the Project $863$ and the National Natural 
Science Foundation of China (59790050, 19874014).


\begin{references}
\bibitem{note5} The Lagrangian for TLM models is described as\cite{wwu,tlm}
$L=\int dx\hat{\psi}^{\dagger}(x) \{i\hbar \partial_t
+i\sigma_3 v_F \partial_x + \sigma_1\Delta \}\hat{\psi}(x)$. The vector
potential $A$ is included by changing 
$-i\hbar\partial_x\to-i\hbar\partial_x-eA$, it does not contain $A^2$ term
directly from the Lagrangian.
\bibitem{note6} Strictly speaking, this gauge phase factor in this paper is 
approximate of the diagonal elements in the matrix of dipole operator 
$\hat{D}$.\cite{gebhard,fradkin} As the example we show above, the diagonal 
term is enough to obtain the convergent results in linear case. But for 
nonlinear cases like $\chi^{(2)}$ and $\chi^{(3)}$, the non-diagonal terms of 
$\hat{D}$ matrix will be important to cancel the ZFD problem through $JJ$ 
correlation. Certainly the procedures to include full gauge phase factor will 
be complicated and thus impractical for $JJ$ correlation.
\end{references}
\end{document}